\documentclass[twoside]{dis09}
\usepackage[latin1]{inputenc}
\usepackage[dvips]{graphicx,epsfig,color}
\usepackage{wrapfig,rotating}
\usepackage{amssymb,amsmath,array}

\newcommand{\mrf}[1]{\mbox{$\mathrm{#1}$}}
\newcommand{\mif}[1]{\mbox{$\mathit{#1}$}}

\newcommand{\DGG}{\mif{\Delta G/G}}

\newcommand{\gom}{\mrf{GeV/\mif{c}}}
\begin{document}
\title{Azimuthal asymmetries in SIDIS off unpolarized targets at COMPASS}

%***********************************************************************
% AUTHORS INFORMATION AREA
%***********************************************************************
\author{Andrea Bressan$^1$ \\ {\small (on behalf of the COMPASS collaboration)}
%
% Optional short acknowledgment: remove next line if non-needed
%\thanks{This is an optional funding source acknowledgment.}
%
% DO NOT MODIFY THE FOLLOWING '\vspace' ARGUMENT
\vspace{.3cm}\\
%
% Addresses and institutions (remove "1- " in case of a single institution)
1- Department of Physics, University of Trieste and Trieste Section of INFN,\\
via Valerio 2, 34127, Trieste, Italy
}
%***********************************************************************
% END OF AUTHORS INFORMATION AREA
%***********************************************************************

\maketitle

\begin{abstract}
Azimuthal asymmetries measured in unpolarized semi-inclusive deep inelastic
scattering bring important information on the inner structure of the nucleons, and
can be used both to estimate the average quark transverse momentum $k_\perp$
and to access the so-far unmeasured Boer-Mulders functions. COMPASS results using
part of the 2004 data collected with a $^6$LiD target and a 160 GeV $\mu^+$ beam are
presented separately for positive and negative hadrons.
\end{abstract}

\section{Introduction}

After years of study to understand how the nucleon spin originates
from the constituent partons, an exhaustive answer is still missing. Moreover, while a lot of
information has been gathered concerning the longitudinal structure  of a fast
moving nucleon (with respect
to its direction of motion), very little is known about the transverse structure. In recent
years these aspects have raised a lot 
of interest and after important theoretical developments and experimental
findings, transverse spin and transverse momentum $k_\perp$ of the quarks are by now
considered 
as fundamental ingredients in the description of the hadron structure.
Spin-$k_\perp$ correlations give rise to various observables in hard hadronic
processes such as the azimuthal asymmetries seen both in unpolarized or
transversely polarized semi-inclusive deep-inelastic scattering (SIDIS), and led
to the introduction of transverse momentum dependent (TMD) parton distribution
functions (PDF) and fragmentation functions (FF). Among these functions of special interest
are the Sivers functions $f^\perp_{1T} (x,k_\perp)$~\cite{sivers}, which describe an azimuthal
asymmetry in the parton distributions inside a transversely polarized nucleon, 
and by their chirally-odd partner $h^\perp_1 (x,k_\perp)$, the Boer-Mulders
functions~\cite{boer}, describing the transverse parton polarization inside an unpolarized
hadron, and generating azimuthal asymmetries in unpolarized SIDIS. 
COMPASS results on the Sivers asymmetries are given
elsewhere~\cite{bressanSSA} while here the unpolarized azimuthal asymmetries are
presented. 

The cross-section for hadron production in lepton-nucleon DIS $\ell N
\rightarrow \ell' h X$ for unpolarized targets and an unpolarized or
longitudinally polarized beam is the following~\cite{bacchetta}:
%\[
%\begin{array}{rcl}
%\frac{d\sigma}{dx dy d\phy_S dz d\phi_h dp^2_{h,T}} &=& \frac{\alpha^2}{xyQ^2}
%\frac{y^2}{2(1-y)} \left( 1+\frac{\gamma^2}{2x} \right) \left[ F_{UU,T} +
%  F_{UU,L} + \sqrt{2\epsilon(1-\epsilon)} \cos \phi_h F^{\cos \phi_h}_{UU} \right.\\[2ex]
%  & & \left. + \epsilon \cos(2\phi_h) F^{\cos\; 2\phi_h}_{UU} + \lambda_l
%  \sqrt{2\epsilon(1-\epsilon)} \sin \phi_h F^{\sin \phi_h}_{LU} \right] 
%\end{array}
%\]
\[
\begin{array}{rcl}
\dfrac{d\sigma}{dx dy dz d\phi_h dp^2_{h,T}} &=&  \dfrac{\alpha^2}{xyQ^2}
\dfrac{1+(1-y)^2}{2} \left[ F_{UU,T} +
  F_{UU,L} + \varepsilon_1 \cos \phi_h F^{\cos \phi_h}_{UU} \right.\\[2ex]
  & & \left. + \varepsilon_2 \cos(2\phi_h) F^{\cos\; 2\phi_h}_{UU} + \lambda_\mu
  \varepsilon_3
  \sin \phi_h F^{\sin \phi_h}_{LU} \right] 
\end{array}
\]
where $\alpha$ is the fine structure constant, $x$, $y$ and $Q^2$ are the
inclusive DIS variables, $z$ is the fraction of the virtual
photon energy carried by the detected hadron, $\phi_h$ is the
azimuthal angle of the outgoing hadron in the $\gamma$-nucleon
system. $F_{UU,T}$,  $F_{UU,L}$, $F^{\cos \phi_h}_{UU}$,  $F^{\cos\;
  2\phi_h}_{UU}$ and $F^{\sin \phi_h}_{LU}$ are structure functions, with the
first and second subscripts which indicate the beam and target polarization
respectively, and the last subscript which, if present, indicates the
polarization of the virtual photon. Finally $\lambda_\mu$ is the beam
longitudinal polarization and: 
\[
\begin{array}{rcl}
\varepsilon_1 & = & \dfrac{2(2-y)\sqrt{1-y}}{1+(1-y)^2} \\[2ex]
\varepsilon_2 & = & \dfrac{2(1-y)}{1+(1-y)^2} \\[2ex]
\varepsilon_3 & = & \dfrac{2 y \sqrt{1-y}}{1+(1-y)^2}
\end{array}
\]
are depolarization factors. The Boer-Mulders PDFs contribute to both
the $\cos \phi$ and the $\cos 2\phi$ structure functions, together with the so
called Cahn effect~\cite{cahn} which arises from the fact that the kinematics is
non collinear when
the $k_\perp$ is taken into account (i.e. a kinematical higher twist), and with the
perturbative gluon radiation, resulting in order $\alpha_s$ QCD processes. pQCD
effects are becoming important for high transverse momenta $p_T$ of the produced hadrons,
while are small for $p_T$ up to 1 GeV$/c$. The $\sin \phi_h$ modulation, which
arises from the natural polarization of the muon beam, does not have a clear
interpretation in the parton model. 

In the past, azimuthal asymmetries have been measured by the EMC
collaboration~\cite{emc1,emc2}, with a liquid hydrogen target and a muon beam at
a slightly higher energy, but without separating hadrons of different
charge. These data have been used~\cite{anselmino} to extract the average
$\langle k_\perp^2 \rangle$. Azimuthal asymmetries have been also measured by E665~\cite{e665} and at
higher energies by ZEUS~\cite{zeus}.  
More recent are the COMPASS results first presented at~\cite{kafer}, and the
measurements done by HERMES, first shown at~\cite{giordano}.

\section{The COMPASS experiment} 

The COMPASS experiment has been set up at the CERN SPS M2 beam line.  It
combines high rate beams with a modern two stage magnetic
spectrometer\cite{nimcompass}. 

COMPASS has collected data with a 160 GeV positive
muon beam impinging on a polarized solid target. The beam is naturally polarized by the
$\pi -$decay mechanism, and the beam polarization is estimated to be $\sim 80\%$
with a $\pm 5\%$ relative error.  The beam intensity is $2\times 10^8$ muons per spill.

Up to 2006 the experiment has used $^6$LiD as deuteron target because its
favorable dilution factor of $\simeq$0.4, particularly important for the
measurement of \DGG. In 2007 an ammonia NH$_3$ target has been used as proton
target. Polarizations of 50\% and 90\% have been reached, respectively for the
two target materials.

\section{Analysis and Systematic Studies}

The event selection requires standard DIS cuts, i.e. $Q^2 > 1\ (\gom)^2$,
mass of the final hadronic state $W>5\ \mrf{GeV}/c^2$, $0.1 < y < 0.9$, and the
detection of at least one hadron in the final state.  Moreover only events with
a vertex in the forward target cell are used in order to minimize nuclear
interactions and for a better data/Monte Carlo agreement. Finally, for the
detected hadrons it is also required that: 

\begin{wrapfigure}{r}{0.50\columnwidth}
\vspace{-30pt}
\begin{center}
\includegraphics[width=0.49\columnwidth]{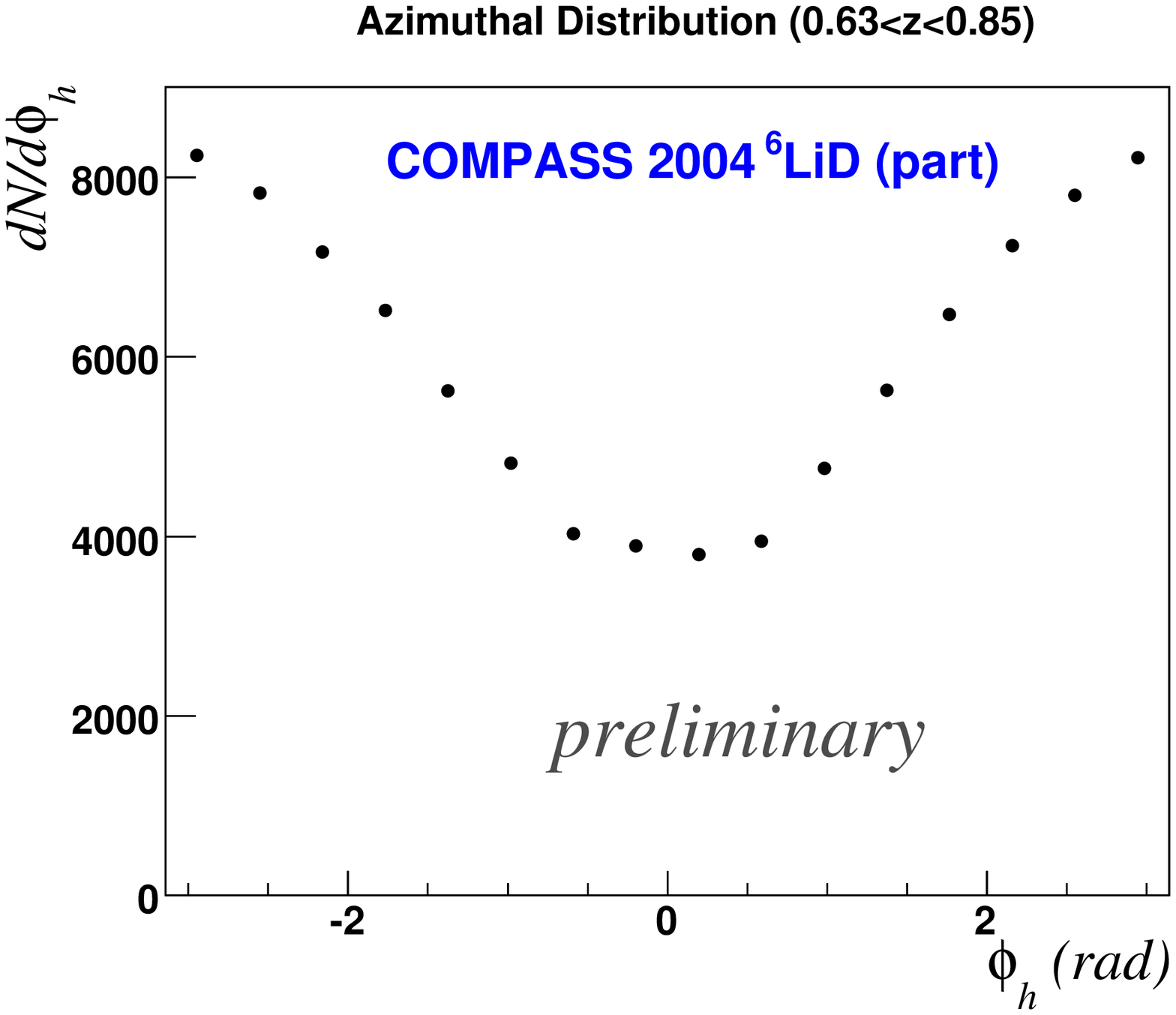}
\includegraphics[width=0.49\columnwidth]{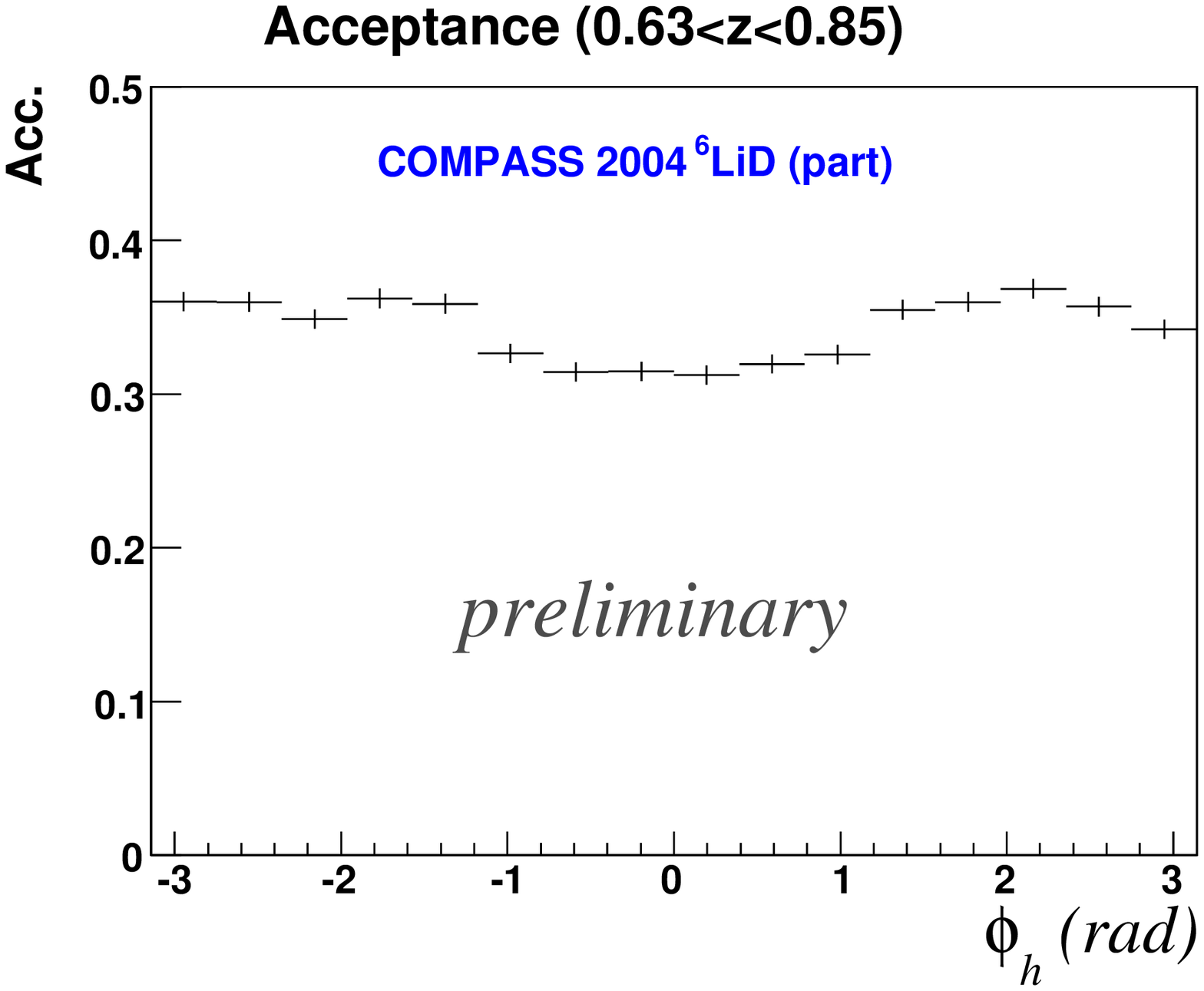}
\includegraphics[width=0.49\columnwidth]{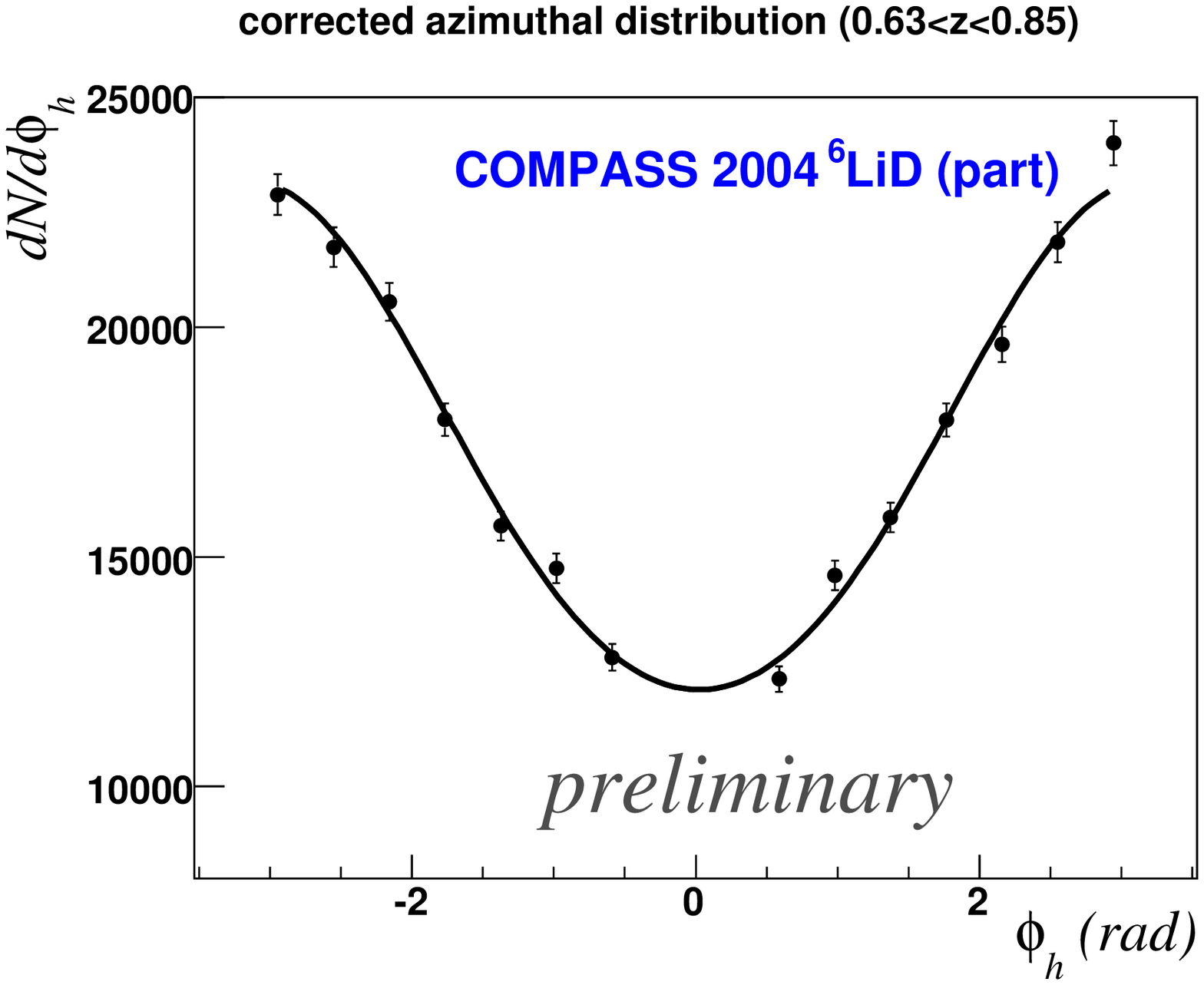}
\end{center}
\vspace{-20pt}
\caption{Top:measured azimuthal distribution after weighting for the two
target polarization. Middle: acceptance distribution as calculated by the Monte
Carlo data. Bottom: azimuthal distribution corrected for the acceptance effects. 
}
\label{azidist}
\end{wrapfigure}

\begin{itemize}
\item the fraction of the virtual photon energy carried by the hadron be
$0.2 < z=E_h/E_\gamma < 0.85$ to select hadrons from the current fragmentation
  region; 
\item
$0.1 < p_T < 1.5\ \mrf{GeV}/c$, where $p_T$ is the hadron transverse momentum with
respect to the virtual photon direction, for a better determination of the
azimuthal angle $\phi_h$. 
\end{itemize}

Data taken both with a longitudinally polarized and a transversely polarized
target have been used, mixing positive or negative orientations in order to
cancel effects depending on the polarization of the nucleons. This statistics
corresponds to almost 1 month of the 2004 data taking and after all
the cuts consists of $5\times 10^6$ positive hadrons and $4\times 10^6$ negative
hadrons entering the asymmetry calculations.

In the measurement of unpolarized asymmetries all the methods used to cancel the
experimental acceptance cannot be adopted and the correction for this effect is
mandatory before fitting the azimuthal modulation. This is done by using a full
Monte Carlo chain, which starts from the SIDIS event generation performed by
Lepto\cite{lepto}, simulate the experimental setup and the particle interactions
in the passive and active material of the detectors (also including the detector
response done by COMGEANT), and ends with the
reconstruction of the generated events by the same 
program (CORAL) used to analyze the real data. The quality of this chain is evaluated by
comparing distributions of real data and of generated events both for the DIS variables and for the
hadronic variables. 

The experimental acceptance as a function of the azimuthal angle $A(\phi)$ is
then calculated as the ratio of 
reconstructed over generated events for each bin of $x$, $z$ and $p_T$ on which
the asymmetries are measured. The overall $y$, $z$ and $p_T$
acceptances are quite constant over the range used in the analysis, so that the
effect coming by the integration over the unlooked variables when the
asymmetries in one of the variables are extracted is well within the systematic error. 
The measured distribution, corrected for acceptance is fitted with the following
functional form:
\[
\begin{array}{lll}
N(\phi) &=&N_0 \left( 1 + A^D_{\cos \phi}  \cos \phi + \right.  \\&& \left. A^D_{\cos 2\phi} \cos 2\phi
 +  A^D_{\sin \phi} \sin \phi \right) 
\end{array}
\]
%\[
%N(\phi) =N_0 \left( 1 + A^D_{\cos \phi}  \cos \phi +  A^D_{\cos 2\phi} \cos 2\phi
% +  A^D_{\sin \phi} \sin \phi \right) 
%\]
An example of a measured azimuthal distribution, acceptance corrections and
corrected azimuthal distribution is shown in Fig.~\ref{azidist}, together with
the resulting fit.

The contribution of the acceptance corrections to the systematic error has been
studied with care.
As the
asymmetries were extracted from data taken both with longitudinal and
transverse target configurations,  comparing the two results gives the effect of the acceptance
changes due to the different configuration (solenoid vs. dipole) of the target
magnet and to the different direction of the incoming beam (for the
transverse setup the beam is bent in order to leave the target with the same direction
as in the longitudinal case). In order to check the effect of the simulation
parameters the acceptances have been calculated using two different sets
of Lepto parameters. All the
resulting asymmetries were compared in order to quantify the
systematic error in each kinematical bin. Further systematic tests,
like splitting the data sample according to the event
topology and to the time of the measurement, gave no significant contributions.
\section{Results and Comments}

\begin{wrapfigure}{r}{0.65\columnwidth}
\vspace{-25pt}
\begin{center}
\includegraphics[width=0.63\columnwidth]{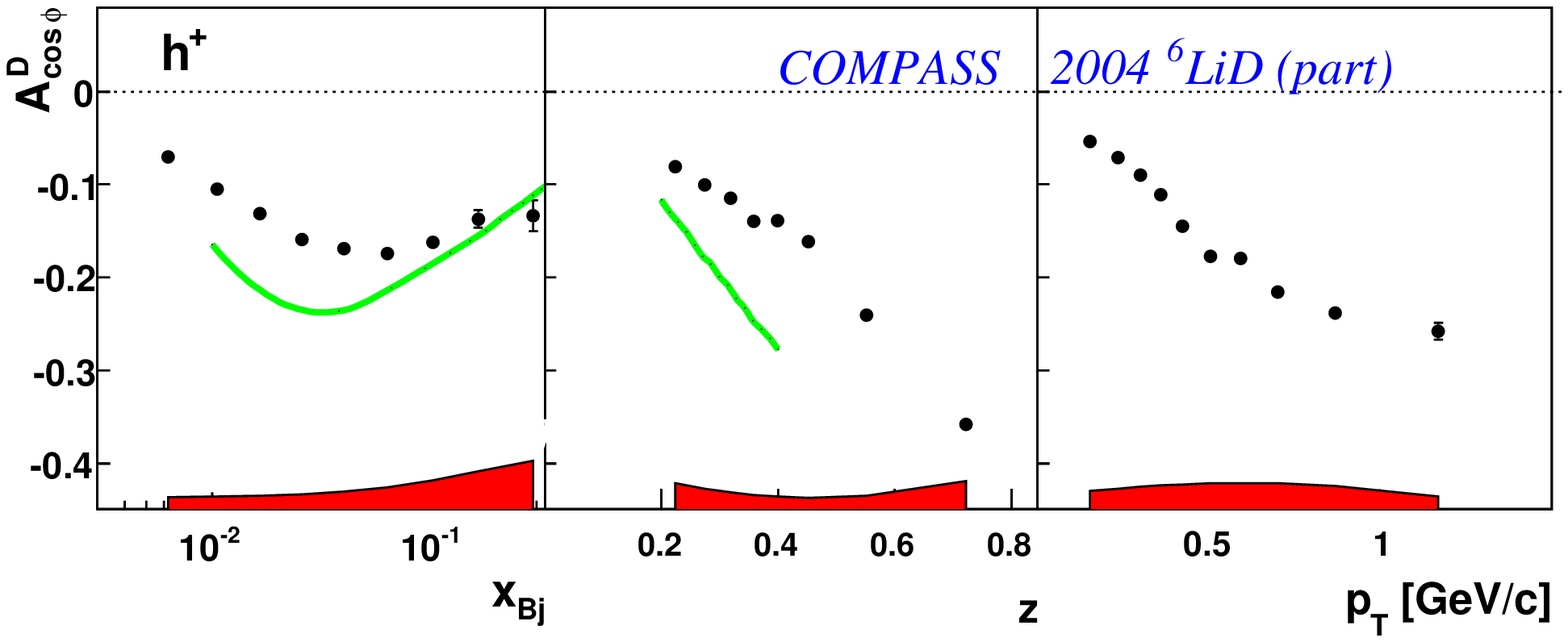}
\includegraphics[width=0.63\columnwidth]{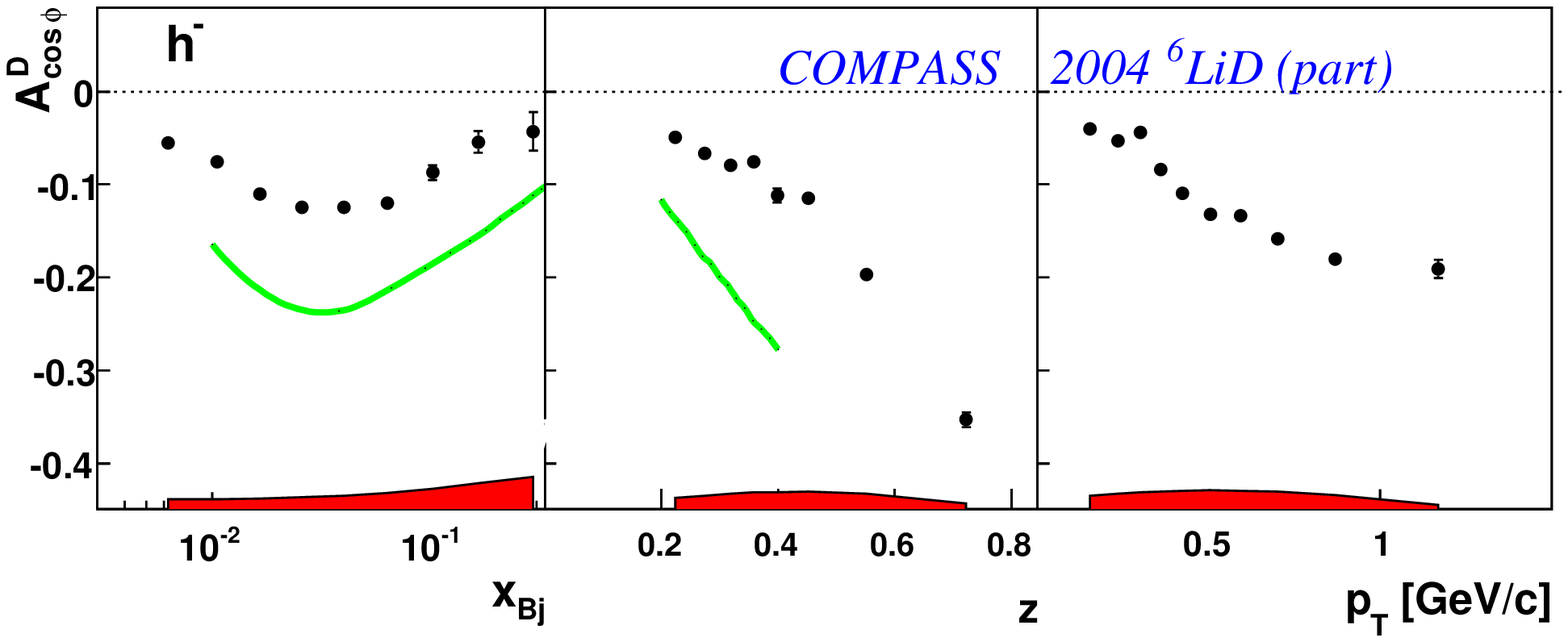}
\end{center}
\vspace{-15pt}
\caption{$\cos \phi$ asymmetries from COMPASS deuteron data
for positive (upper row) and negative (lower
raw) hadrons; the asymmetries includes the kinematical factor
$\varepsilon_1$ and the bands indicate the size of the systematic errors. 
The superimposed curves are the values predicted by~\protect\cite{anselmino2}
taking into account the Cahn effect only.
}
\label{f:cosphi}
\end{wrapfigure}

The $\sin \phi$ asymmetries, not shown here, measured by COMPASS are compatible
with zero, at the present level of statistical and systematic errors, over the
full range of $x$, $z$ and $p_T$ covered by the data.

The $\cos \phi$ asymmetries extracted from COMPASS deu\-teron data
are shown in Fig.~\ref{f:cosphi} for positive (upper row) and negative (lower
row) hadrons, as a function of $x$, $z$ and $p_T$. The bands indicate the size
of the systematic error. The asymmetries show the same trend for positive and
negative hadrons with a slightly larger values for the positive one. Values as
large as 30$\div$40\% are reached in the last point of the $z$ range. The theoretical
predictions~\cite{anselmino2} in Fig.~\ref{f:cosphi} takes into account the Cahn
effect only, which
does not depend on the hadron charge. The Boer-Mulders PDFs are not
taken into account in this case. 

\begin{wrapfigure}{r}{0.6\columnwidth}
\vspace{-20pt}
\begin{center}
\includegraphics[width=0.5\textwidth]{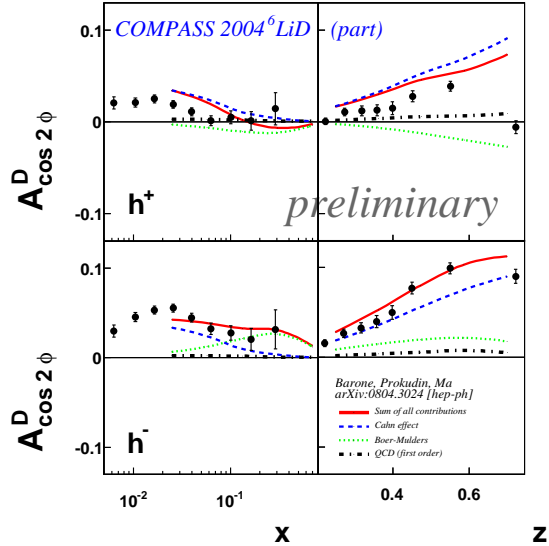}
\end{center}
\vspace{-20pt}
\caption{$\cos 2 \phi$ asymmetries from COMPASS deuteron data
for positive (upper row) and negative (lower
raw) hadrons; the asymmetries are divided by the kinematical factor
$\varepsilon_1$ and the red bands indicate size of the systematic errors. 
}
\label{f:cos2phi}
\end{wrapfigure}

The $\cos 2 \phi$ asymmetries are shown in Fig.~\ref{f:cos2phi} together with
the theoretical predictions of~\cite{barone}, which take into account
the kinematical contribution given by the Cahn effect, first order pQCD (which,
as expected, is negligible in the low $p_T$ region), and 
the Boer-Mulders PDFs (coupled to the Collins FF), which give a different
contribution to positive and negative 
hadrons. In~\cite{barone} the Boer-Mulders PDFs are assumed to be proportional
to the Sivers function as extracted from the preliminary HERMES data. The
COMPASS data show different amplitude 
for positive and negative hadrons, a trend which confirms the theoretical
predictions. There is a satisfactory agreement between the data points and the
model calculations, which hints to a non zero Boer-Mulders PDFs.

% ****************************************************************************
% BIBLIOGRAPHY AREA
% ****************************************************************************

\begin{footnotesize}
% IF YOU DO NOT USE BIBTEX, USE THE FOLLOWING SAMPLE SCHEME FOR THE REFERENCES
% ----------------------------------------------------------------------------

\end{footnotesize}

% ****************************************************************************
% END OF BIBLIOGRAPHY AREA
% ****************************************************************************


\begin{thebibliography}{99}
% Please replace the numbers for   contribId   and   sessionId
% in the following URL. You can get this information by going to 
% http://indico.cern.ch/confAuthorIndex.py?confId=24657
% and search for your contribution and click on the title
% Be aware: '&amp;' must be replaced by simple '&' as in example below
\bibitem{url} Slides: \\ 
\verb$http://indico.cern.ch/contributionDisplay.py?contribId=308&sessionId=4&confId=53294$
%------- replace following references ;-)
%\bibitem{parton_qed} A.D.~Martin {\it et~al.}, Eur. Phys. J. {\bf C39} 155 (2005).
%\bibitem{H1}N.~Gogitidze, arXiv:hep-ex/0701033 (2007).
%\bibitem{DVCS}S.~Friot, B.~Pire and L.~Szymanowski, Phys. Lett. {\bf B645} 153 (2007);\\
%              D.~Hasell, R.~Milner and K.~Takase, AIP Conf. Proc. {\bf 588} 187 (2001);\\
%              M.~Krawczyk and A.~Zembrzuski, Phys. Rev. {\bf D57} 10 (1998).
%\bibitem{pomeron}R.~Brower and C.~Tan, PoS LAT2005 279 (2006);\\
%                 J.P.~Guillaud and A.~Sobol,
%  {\it Perspectives of the study of double Pomeron exchange at the LHC},
%  11th Lomonosov Conference on Elementary Particle Physics, Moscow, Russia (2003).
% ----------------------------------------------------------------------------
\bibitem{sivers}
D.W. Sivers, Phys.\ Rev.\ {\bf D41} (1991) 83.
\bibitem{boer}
D. Boer and P.J. Mulders, Phys. Rev. {\bf D57} (1998) 5780.
\bibitem{bressanSSA}
A. Bressan, ``COMPASS Results on Collins and Sivers Asymmetries'', these
proceedings.
\bibitem{bacchetta}
A. Bacchetta {\it et al.}, JHEP {\bf 0702} (2007) 93.
\bibitem{cahn}
R.N. Cahn, Phys. Lett. {\bf B78} (1978) 269.
\bibitem{emc1}
The EMC Collaboration, M. Arneodo {\it et al.}, Z. Phys. {\bf C34} (1987) 277.
\bibitem{emc2}
The EMC Collaboration, J. Ashman {\it et al.}, Z. Phys. {\bf C52} (1991) 361.
\bibitem{anselmino}
M. Anselmino {\it et al.}, Phys. Rev. {\bf D71} (2005) 74006
\bibitem{e665}
The E665 Collaboration, M.R. Adams {\it et al.}, Phys. Rev. {\bf D48} (1993)
5057.
\bibitem{zeus}
The ZEUS Collaboration, J. Breitweg {\it et al.}, Phys. Lett. {\bf B481} (2000) 199.
\bibitem{kafer}
W. K\"afer, ``Measurements of unpolarized azimuthal asymmetries at COMPASS'', in
the proceedings of `Transversity 2008' and arXiv:0808.0140.
\bibitem{giordano}
F. Giordano, ``Measurements of azimuthal asymmetries of the unpolarized
cross-section at HERMES'', in the proceedings of 'SPIN2008' and arXiv:0901.2438.
\bibitem{nimcompass}
The COMPASS Collaboration, P.~Abbon {\it et al.}, Nucl. Instrum. Meth. {\bf
  A577} (2007) 455.
\bibitem{lepto}
G. Ingelman, A. Edin and J. Rathsman, Comp. Phys. Commun. 101 (1997) 108.
\bibitem{anselmino2}
M.~Anselmino {\it et al.}, Eur.\ Phys.\ J.\  {\bf A31}
(2007) 373.
\bibitem{barone}
V. Barone, A. Prokudin and B.Q. Ma, Phys. Rev. {\bf D78} (2008) 45022.
\end{thebibliography}
\end{document}